\documentclass[]{spie}  

\usepackage{amsmath,amsfonts,amssymb,textcomp}
\usepackage{xcolor}
\usepackage{graphicx}
\usepackage[colorlinks=true, allcolors=blue]{hyperref}

\title{Verification of the Optical System of the 9.7-m Prototype Schwarzschild-Couder Telescope}

\author[a]{C.~Adams}
\affil[a]{Physics Department, Columbia University, New York, NY 10027, USA}

\author[b]{R.~Alfaro}
\affil[b]{Instituto de F\'{i}sica, Universidad Nacional Aut\'{o}noma de M\'{e}xico, Ciudad de Mexico, Mexico}

\author[c]{G.~Ambrosi} 
\affil[c]{INFN Sezione di Perugia, Perugia, Italy}

\author[d]{M.~Ambrosio}
\affil[d]{INFN Napoli, Italy}

\author[d]{C.~Aramo}
\author[e]{W.~Benbow}
\affil[e]{Center for Astrophysics | Harvard \& Smithsonian, Cambridge, MA 02138, USA}

\author[c,f]{B.~Bertucci}
\affil[f]{Universit\`a degli Studi di Perugia, Perugia, Italy}

\author[g,h]{E.~Bissaldi}
\affil[g]{Dipartimento Interateneo di Fisica dell'Universit\`a e del Politecnico di Bari}
\affil[h]{INFN Bari, Via E. Orabona 4, 70125 Bari, Italy}

\author[h]{M.~Bitossi}
\author[d]{A.~Boiano}
\author[d]{C.~Bonavolont\`a}
\author[i]{R.~Bose}
\affil[i]{Department of Physics, Washington University, St. Louis, MO 63130, USA}

\author[a]{A.~Brill}
\author[i]{J.~H.~Buckley}
\author[j]{K.~Byrum}
\affil[j]{Argonne National Laboratory, 9700 S. Cass Avenue, Argonne, IL 60439, USA}

\author[k]{R.~A.~Cameron}
\affil[k]{Kavli Institute for Particle Astrophysics and Cosmology, SLAC National Accelerator Laboratory, Stanford University, Stanford, CA 94305, USA}

\author[n]{M.~Capasso}
\author[c]{M.~Caprai}
\author[l]{C.~E.~Covault}
\affil[l]{Department of Physics, Case Western Reserve University, Cleveland, Ohio 44106, USA}

\author[h]{L.~Di~Venere}
\author[m]{S.~Fegan}
\affil[m]{LLR/Ecole Polytechnique, Route de Saclay, 91128 Palaiseau Cedex, France }

\author[n]{Q.~Feng$^{\dagger}$}
\affil[n]{Department of Physics and Astronomy, Barnard College, Columbia University, NY 10027, USA}

\author[c,f]{E.~Fiandrini}
\author[o]{A.~Furniss}
\affil[o]{Department of Physics, California State University - East Bay, Hayward, CA 94542, USA}

\author[p]{M.~Garczarczyk}
\affil[p]{Deutsches Elektronen Synchrotron (DESY), Platanenallee 6, D-15738 Zeuthen, Germany}

\author[q]{F.~Garfias}
\affil[q]{Instituto de Astronom\'ia, Universidad Nacional Aut\'onoma de M\'exico, Ciudad de M\'exico, Mexico}

\author[r]{A.~Gent}
\affil[r]{School of Physics \& Center for Relativistic Astrophysics, Georgia Institute of Technology, 837 State Street NW, Atlanta, GA 30332-0430, USA}
\author[g,h]{N.~Giglietto}
\author[g,h]{F.~Giordano}
\author[q]{M.~M.~Gonz\'alez}
\author[l]{R.~Halliday}
\author[s]{O.~Hervet}
\affil[s]{Santa Cruz Institute for Particle Physics and Department of Physics, University of California, Santa Cruz, CA 95064, USA}

\author[t,u]{G.~Hughes}
\affil[t]{Fred Lawrence Whipple Observatory, Harvard-Smithsonian Center for Astrophysics, Amado, AZ 85645, USA}
\affil[u]{CTAO, Saupfercheckweg 1, 69117 Heidelberg, Germany}

\author[a]{T.~B.~Humensky}
\author[c]{M.~Ionica}
\author[q]{A.~Iriarte} 

\author[v]{W.~Jin}
\affil[v]{Department of Physics and Astronomy, University of Alabama, Tuscaloosa, AL 35487, USA}

\author[w]{P.~Kaaret}
\affil[w]{Department of Physics and Astronomy, University of Iowa, Iowa City, IA 52242, USA}

\author[x]{D.~Kieda}
\affil[x]{Department of Physics and Astronomy, University of Utah, Salt Lake City, UT 84112, USA}

\author[y]{B.~Kim}
\affil[y]{Department of Physics and Astronomy, University of California, Los Angeles, CA 90095, USA}

\author[h]{F.~Licciulli}

\author[z]{M.~Limon}
\affil[z]{Department of Physics and Astronomy, 
University of Pennsylvania, Philadelphia, PA 19104, USA}

\author[g,h]{S.~Loporchio}

\author[d]{V.~Masone}

\author[aa]{T.~Meures}
\affil[aa]{Department of Physics and Wisconsin IceCube Particle Astrophysics Center, University of Wisconsin, Madison, WI 53706, USA}

\author[aa]{B.~A.~W.~Mode}

\author[n]{R.~Mukherjee}

\author[ab]{D.~Nieto}
\affil[ab]{Instituto de Física de Partículas y del Cosmos, Universidad Complutense de Madrid, Spain}

\author[ac]{A.~Okumura}
\affil[ac]{Institute for Space--Earth Environmental Research and Kobayashi--Maskawa Institute for the Origin of Particles and the Universe, Nagoya University, Nagoya 464-8601, Japan}

\author[r]{N.~Otte}

\author[ad]{N.~Palombara}
\affil[ad]{INAF - IASF Milano, Italy}

\author[g,h]{F.~R.~Pantaleo}
\author[ai,ae]{R.~Paoletti}
\affil[ae]{Dipartimento di Scienze Fisiche, della Terra e dell'Ambiente, Universit\`a degli Studi di Siena, Siena, Italy}

\author[af]{G.~Pareschi}
\affil[af]{INAF - Osservatorio Astronomico di Brera, Milano/Merate, Italy}

\author[ag]{A.~Petrashyk}
\affil[ag]{Citadel Securities LLC, Chicago, IL 60603, USA}
\author[v]{J.~Powell}
\author[r]{K.~Powell}
\author[a]{D.~Ribeiro}
\author[t]{E.~Roache}
\author[ah]{J.~Rousselle}
\affil[ah]{Subaru Telescope NAOJ, Hilo HI 96720, USA}

\author[ai]{A.~Rugliancich}
\affil[ai]{INFN Sezione di Pisa, Pisa, Italy}

\author[q]{J.~Ru\'{i}z-D\'{i}az-Soto}
\author[v]{M.~Santander}
\author[p,u]{S.~Schlenstedt}
\author[ad]{S.~Scuderi}

\author[y]{R.~Shang}
\author[af]{G.~Sironi}
\author[y]{B.~Stevenson}
\author[ai,ae]{L.~Stiaccini}
\author[aa]{L.~P.~Taylor}
\author[c,f]{L.~Tosti}
\author[q]{G.~Tovmassian}
\author[c,f,aj]{V.~Vagelli}
\affil[aj]{ASI Italian Space Agency - Scientific Research Unit, Roma, 00133, Italy}
\author[ak,d]{M.~Valentino}
\affil[ak]{CNR-ISASI, Italy}
\author[aa]{J.~Vandenbroucke}
\author[y]{V.~V.~Vassiliev$^{\dagger}$}
\author[al]{S.~P.~Wakely}
\affil[al]{Enrico Fermi Institute, University of Chicago, Chicago, IL 60637, USA}
\author[am]{P.~Wilcox}
\affil[am]{School of Physics and Astronomy, University of Minnesota, Minneapolis, MN 55455, USA}
\author[s]{D.~A.~Williams} 
\author[y]{P.~Yu}
\author[ ]{for the CTA SCT Project}

\authorinfo{$^{\dagger}$ E-mails: \href{qifeng@nevis.columbia.edu}{qifeng@nevis.columbia.edu}, \href{vvv@astro.ucla.edu}{vvv@astro.ucla.edu}
}

\begin{document} 

\maketitle

\begin{abstract}
For the first time in the history of  ground-based $\gamma$-ray astronomy, the on-axis performance of the dual mirror, aspheric, aplanatic Schwarzschild-Couder optical system has been demonstrated in a $9.7$-m aperture imaging atmospheric Cherenkov telescope. The novel design of the prototype Schwarzschild-Couder Telescope (pSCT) is motivated by the need of the next-generation Cherenkov Telescope Array (CTA) observatory to have the ability to perform wide ($\geq 8^{\circ}$) field-of-view observations simultaneously with superior imaging of atmospheric cascades (resolution of $0.067^{\circ}$ per pixel or better). The pSCT design, if implemented in the CTA installation, has the potential to improve significantly both the $\gamma$-ray angular resolution and the off-axis sensitivity of the observatory, reaching nearly the theoretical limit of the technique and thereby making a major impact on the CTA observatory sky survey programs, follow-up observations of multi-messenger transients with poorly known initial localization, as well as on the spatially resolved spectroscopic studies of extended $\gamma$-ray sources. This contribution reports on the initial alignment procedures and point-spread-function results for the challenging segmented aspheric primary and secondary mirrors of the pSCT.
\end{abstract}

\keywords{Imaging Cherenkov Telescopes, Aplanatic Optical System, Gamma Ray}

\section{INTRODUCTION}
\label{sec:intro}  

Very high-energy (VHE) $\gamma$-ray astrophysics in the energy domain above $100$ GeV has emerged during the last fifteen years as a branch of astronomical science due to the success of the ground-based $\gamma$-ray detection technology utilizing small arrays ($\leq 5$ telescopes) of large aperture ($\geq 10$ m) imaging atmospheric Cherenkov telescopes (IACTs). IACTs detect VHE photons indirectly by measuring Cherenkov light, which is produced by the cascades of secondary particles resulting from the interaction of the primary photons with the Earth's atmosphere. Cherenkov photons detected by IACTs are broadly distributed in the UV-visible wavelength band between $250$ and $800$ nm, peaking around $350-400$ nm, and they are produced by secondary electrons and positrons at distances of a few to a few tens of kilometers from the IACTs. The typical duration of the Cherenkov light pulse from a single atmospheric cascade ranges from $6$ to $200$ ns depending on the photon energy and the geometrical orientation of the trajectory of the primary photon with respect to the array of IACTs. Given these characteristics of the Cherenkov light, IACTs do not operate in the diffraction-limited regime. The main goals of optically fast IACTs are to collect as many Cherenkov photons as possible during an extremely short exposure, and to capture the image of the atmospheric cascade at as high a resolution as possible. The imaging resolution is an important characteristic of IACT performance, as it directly affects gamma-ray angular resolution and the efficiency of discrimination against the isotropic background of cosmic rays (CRs) impacting the Earth's atmosphere, which dominates IACT triggers. 
 
The optical systems (OSs) of all present-day IACTs are based on a single mirror, prime-focus Davies-Cotton or parabolic design. Although the photons collected by these instruments have a wavelength in the UV and visible range, the alignment requirements for the OSs of these telescopes resemble those of radio telescopes of similar apertures operating at a few tens of mm wavelengths in the diffraction limit. These conventional IACTs, although adequate for all present-day ground-based observatories, such as VERITAS, H.E.S.S. and MAGIC, have become a limiting factor for the improvement of the angular resolution and sensitivity (particularly off-axis) in the context of implementation of the Cherenkov Telescope Array (CTA), the next-generation world-wide observatory currently under design and construction~\citenum{Actis2011CTAconcept,2019scta}. 
The CTA SCT Project team
has built and demonstrated an innovative $9.7$-m dual-mirror prototype Schwarzschild-Couder Telescope (pSCT) for CTA, which has the potential to transform the field of VHE $\gamma$-ray astronomy by offering major improvements in $\gamma$-ray angular resolution and off-axis sensitivity over a wide $8^{\circ}$ field of view (FoV). 
These improvements can potentially lead to a major impact on the key science projects of the CTA observatory \citenum{2019scta}, including the Galactic plane and extragalactic survey programs, follow-up observations of multi-messenger transients often with poorly known initial localization, as well as on the spatially resolved spectroscopic studies of extended $\gamma$-ray sources that may be associated with cosmic ray PeVatrons.

\begin{figure} [ht]
   \begin{center}
   \begin{tabular}{c} 
   \includegraphics[height=10cm]{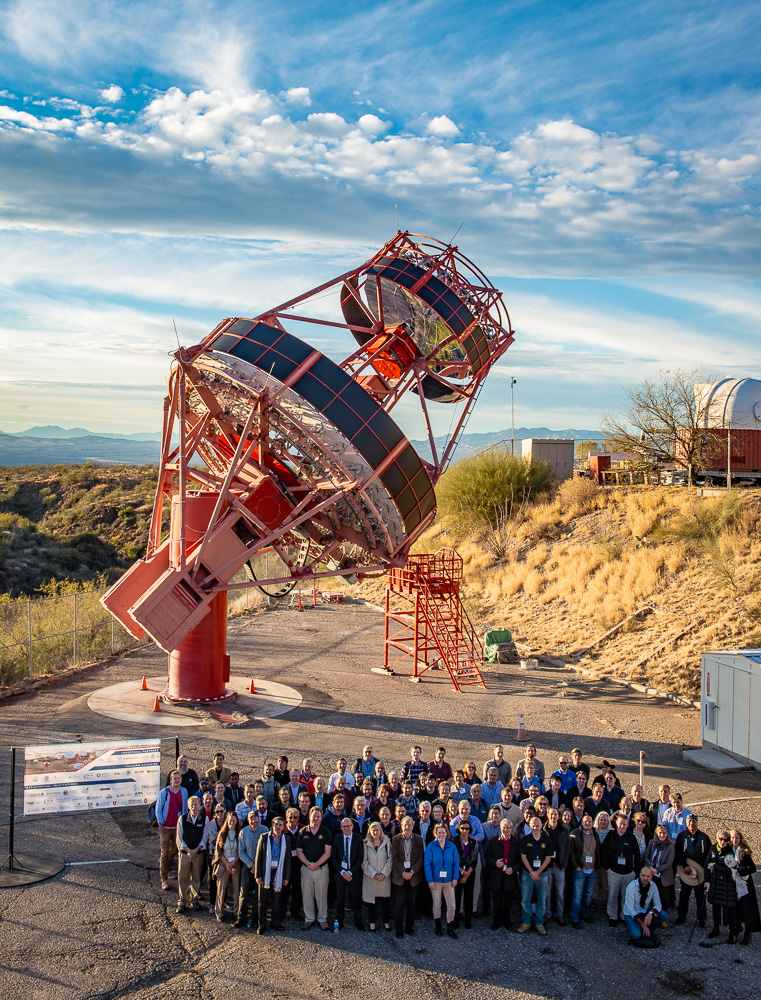}
   \includegraphics[height=10cm]{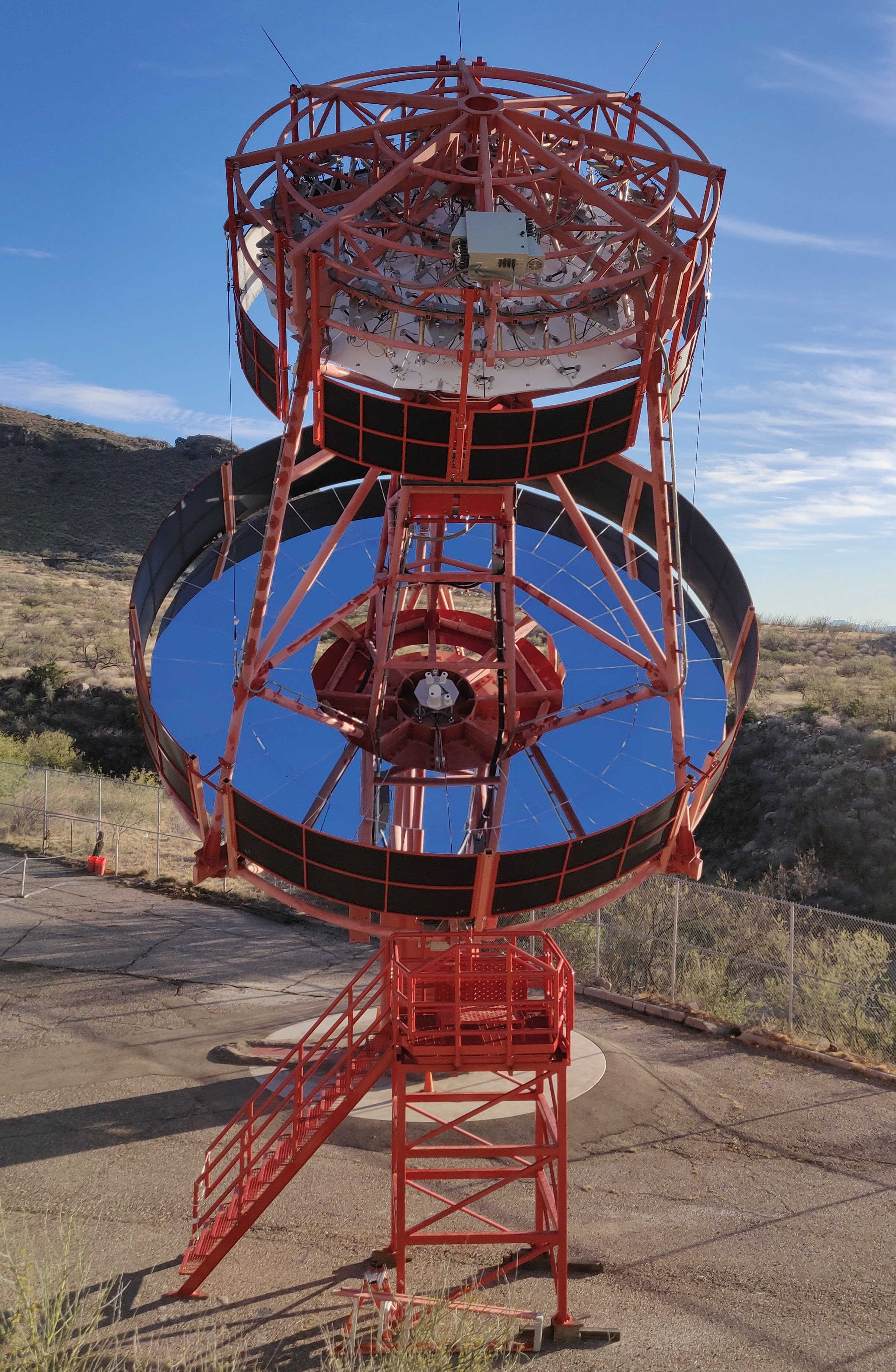}
   \end{tabular}
   \end{center}
   \caption[inaug] 
   { \label{fig:inaug} 
({\it Left}) The inauguration of the pSCT at FLWO on January 17, 2019. 
({\it Right}) A picture of the pSCT showing the segmented primary mirror from the front, segmented secondary mirror from the back, and the camera access tower, which enables the maintenance of the camera positioned between primary and secondary mirrors in the Schwarzschild-Couder OS. }
   \end{figure}

The pSCT incorporates original ideas for the implementation of aplanatic dual-mirror optics, rapid telescope re-positioning mechanics, and a very fast,  modular, and highly integrated design for the camera electronics. The pSCT is delivered at a cost slightly higher but competitive to the conventional designs of IACTs of similar aperture for CTA in spite of the complexity of its optics and at least an order of magnitude more demanding alignment requirements to achieve its superior imaging optical performance in comparison with IACTs of the conventional design. The pSCT has been designed, procured and constructed at the Fred Lawrence Whipple Observatory (FLWO) at the cost of US\$~5.2M funded by NSF (US\$~3.9M) and participating institutions (US\$~1.3M). The pSCT was inaugurated (see Figure~\ref{fig:inaug}) and achieved first light in January 2019. The project is currently undergoing commissioning and is in an early operational phase, which started in June 2019 and lasted until February 2020, when it was interrupted by the development of the COVID-19 pandemic in the U.S. This submission describes the on-axis optical alignment strategy and the results achieved by the project so far.

This submission is organized as follows: an overview of the pSCT OS is given in Section~\ref{sec:overview}, a three-step alignment procedure of the OS is described in Sections \ref{sec:calib}, \ref{sec:p2pas}, and \ref{sec:opt_align}, and the on-axis point-spread-function (PSF) results from the initial commissioning campaign are shown in Section~\ref{sec:psf-res}.

\section{Overview of the pSCT Optical System}
\label{sec:overview}  

The design choices behind the pSCT are based on the desire to simultaneously increase the imaging resolution of atmospheric cascades and the FoV of the Cherenkov camera while maintaining roughly the same telescope cost envelope. Comparing to the conventional Davies-Cotton (DC) Medium-Sized Telescope (MST) of the CTA observatory, which has a $0.17^{\circ}$ diameter pixel, the pSCT imaging pixel size is reduced by a factor of $2.54$ to $0.067^{\circ}$. To populate the desired $8^{\circ}$ FoV, which is approximately twice as large as the FoV of IACTs currently in operation, the MST has a camera with $\sim$1800 pixels (two camera versions exist Flashcam- 1764 pixels~\citenum{Puehlhofer2015FlashCam} \& NectarCam- 1855 pixels~\citenum{Tavernier2019NectarCam}), while a fully instrumented camera for the pSCT would be composed of
11,328 pixels, a factor of 6.3 larger. It is evident that the “same cost envelope” requirement cannot be sustained by scaling the MST design, in which the camera cost is roughly 50\% of the cost of the MST, and dramatic changes in the technology chosen to implement both the OS and the camera are required for the pSCT to achieve this goal. 

The dual-mirror Schwarzschild-Couder (SC) OS of the pSCT is chosen for its ability to fully correct spherical and comatic aberrations over the $8^{\circ}$ FoV and minimize astigmatism by curving the surface of the focal plane. In addition to improving the imaging resolution, the fast SC OS uses a de-magnifying secondary mirror, M2, to reduce the effective focal length of the OS from the $16$ m of the MST to $5.6$ m, thereby reducing the plate scale by a factor of $2.86$ to the value of $1.625$ mm/arcmin. This major reduction of the pSCT camera diameter from $2.23$ m of the MST to $78$ cm enables a change of photosensor technology from traditional photo-multiplier tubes (PMTs) to higher photon detection efficiency silicon photomultipliers (SiPMs) with $6.53 \times 6.53$ mm$^2$ imaging pixels~\citenum{Vassiliev07}. Hence, the plate-scale reduction enables cost-efficient construction of the pSCT camera in which all the front-end and back-end high-density electronics (with 1-ns digitization speed) are located directly behind the photosensors.

The breakthrough of the pSCT in IACT technology, however, comes at the cost of the significantly increased complexity of the SC OS and demanding alignment requirements which approach those of sub-mm radio telescopes of similar aperture. The pSCT OS consists of a $9.7$-m segmented primary mirror (M1: 48 = 32+16 panels), and a $5.4$-m segmented secondary mirror (M2: 24 = 16+8 panels); both are aspheric and M2, in particular, is very curved with a few-cm sag in each mirror panel. The 72 aspheric light-weight hybrid mirror panels have been produced in collaboration with industry by employing cost-effective hot and cold glass slumping replication technologies. A total of 74 microcomputers (72 for mirror panels and 2 for optical tables), 444 actuators (6 actuators for each of 74 Stewart platforms) and 306 mirror panel edge sensors (MPESs) are used to align the pSCT OS. A comprehensive collection of characteristics of the pSCT is shown in Table~\ref{tab:telescope}.

\begin{table}  
\vspace{-2ex}
\caption{\label{tab:telescope} Characteristics of the prototype Schwarzschild-Couder Telescope.}
\vspace{-1ex}
\small
\begin{center}
\begin{tabular}{|l|c|c|}
\hline
Quantity & Value/Summary & Units \\ \hline
Main and Effective Aperture Size & 9.66 / 7.82  & m \\ \hline
System Effective Focal Length & 5.59 & m \\ \hline
Separation between Primary and Secondary Mirrors & 8.39 & m \\ \hline
Separation between Secondary Mirror and Focal Plane & 1.86 & m \\ \hline
Primary Mirror (M1) Radius Max \& Min & 4.83 / 2.19 & m \\ \hline
Number of Mirror Segments in M1 & 48 = 16 (P1) + 32 (P2) & \\ \hline
Average Area of M1 Mirror Segment & 1.2 & m$^{2}$ \\ \hline
Average Weight of M1 Hybrid Mirror Segment & 24 & kg \\ \hline
Weight of M1 Mirror Segment Positioning System & 19 & kg \\ \hline
Secondary Mirror (M2) Radius Max \& Min & 2.71 / 0.40 & m \\ \hline
Number of Mirror Segments in M2 & 24 = 8 (S1) + 16 (S2) & \\ \hline
Average Area of M2 Mirror Segment & 0.94 & m$^{2}$ \\ \hline
Average Weight of M2 Hybrid Mirror Segment & 19 & kg \\ \hline
Weight of M2 Mirror Segment Positioning System & 18 & kg \\ \hline
Total Light Collecting Area on-axis \& at 4 deg off-axis & 50.31 / 47.73 & m$^{2}$ \\ \hline
Vignetting at the FoV Edge & -5.17 & \% \\ \hline
Wavelength Range of Cherenkov Photons & 250 - 800 & nm \\ \hline
M1 Mirror Segment Replication Technology & Cold Glass Slumping & \\ \hline
M2 Mirror Segment Replication Technology & Hot/Cold Glass Slumping & \\ \hline
M1 \& M2 Mirror Surface Vacuum Deposition Coating & Al/Cr/SiO$_2$/HfO$_2$/SiO$_2$ & \\ \hline
Mirror Segment Actuators Precision and Range & 0.003 / 51 & mm \\ \hline
Mirror Segment Positioning Degrees of Freedom & 6 & \\ \hline
Telescope total Moving Mass & 30 & ton \\ \hline
M1 OSS with Mirrors/Baffles & 6.5 & ton \\ \hline
M2 OSS with Mirrors/Baffles & 3.5 & ton \\ \hline
Camera \& Camera Support Structure & 6 & ton \\ \hline
Counterweights & 14 & ton \\ \hline
Telescope Mass and Type of Material for Support Structure & 75 (steel) & ton\\ \hline
Type of mount used for pointing & Alt- azimuth & \\ \hline
Azimuth allowed range & [-270, +270] & deg \\ \hline
Elevation allowed range & [-5, +91] & deg \\ \hline
Camera Diameter & 0.78 & m \\ \hline
Camera Field of View (FoV) Diameter & 8 & deg \\ \hline
Camera FoV solid angle & 50.35 & deg$^2$ \\ \hline
Focal Plane Figure & Parabolic &  \\ \hline
Focal Plane Sag at the FoV Edge & -22 & mm \\ \hline
Focal Plane Plate Scale & 1.625 & mm/arcmin \\ \hline
Characteristic photon incidence angle & 51.25 & deg \\ \hline
PSF at the FoV edge (2 $\times$ MAX[RMS]) & 3.8-4.5 & arcmin \\ \hline
\end{tabular}
\vspace{-3ex}
\end{center}
\end{table}

\begin{figure} [ht]
\begin{center}
\begin{tabular}{c} 
\includegraphics[height=8cm]{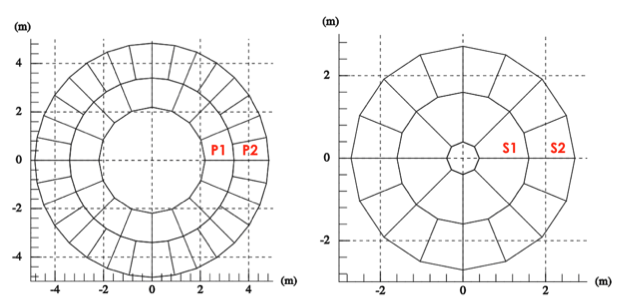}
\end{tabular}
\end{center}
\caption[GAS] 
{ \label{fig:mirror_segmentation} 
The primary mirror, with a radius of 4.83 m (left diagram), is segmented into 48 panels; each inner-ring panel (P1) has an area of 1.33 $\text{m}^{2}$ and each outer-ring panel (P2) has an area of 1.16 $\text{m}^{2}$.
The secondary mirror, with a radius of 2.71 m (right diagram), is segmented into 24 panels; each inner-ring panel (S1) has an area of 0.94 $\text{m}^{2}$ and each outer-ring panel (S2) also has an area of 0.94 $\text{m}^{2}$. Figures are taken from \citenum{Rousselle2015}. 
}
\end{figure}

The unique feature of the pSCT is that both the primary and, particularly, the secondary mirrors are segmented for cost reduction. Both M1 and M2 consist of two types of segments, as explained in Figure \ref{fig:mirror_segmentation}, which are denoted P1 and P2 for the inner and outer primary panels, respectively, and S1 and S2 for the inner and outer secondary panels, respectively. The selected segmentation scheme is optimized for cost, achieving a balance between the fabrication cost of large mirror panels and the cost associated with the complexity of the alignment system. 

\begin{figure} [ht]
\begin{center}
\begin{tabular}{c} 
\includegraphics[height=6cm]{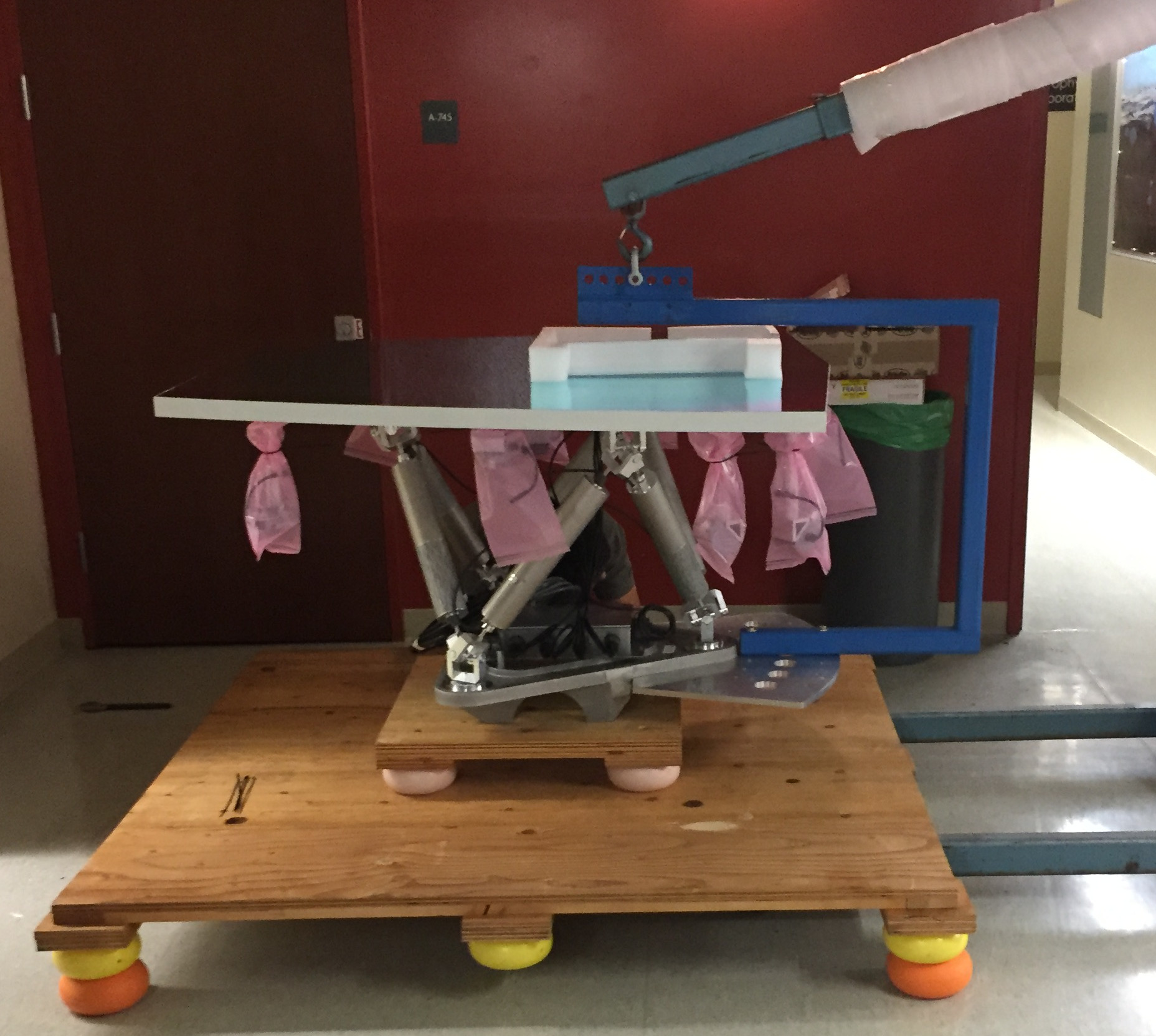}
\includegraphics[height=6cm]{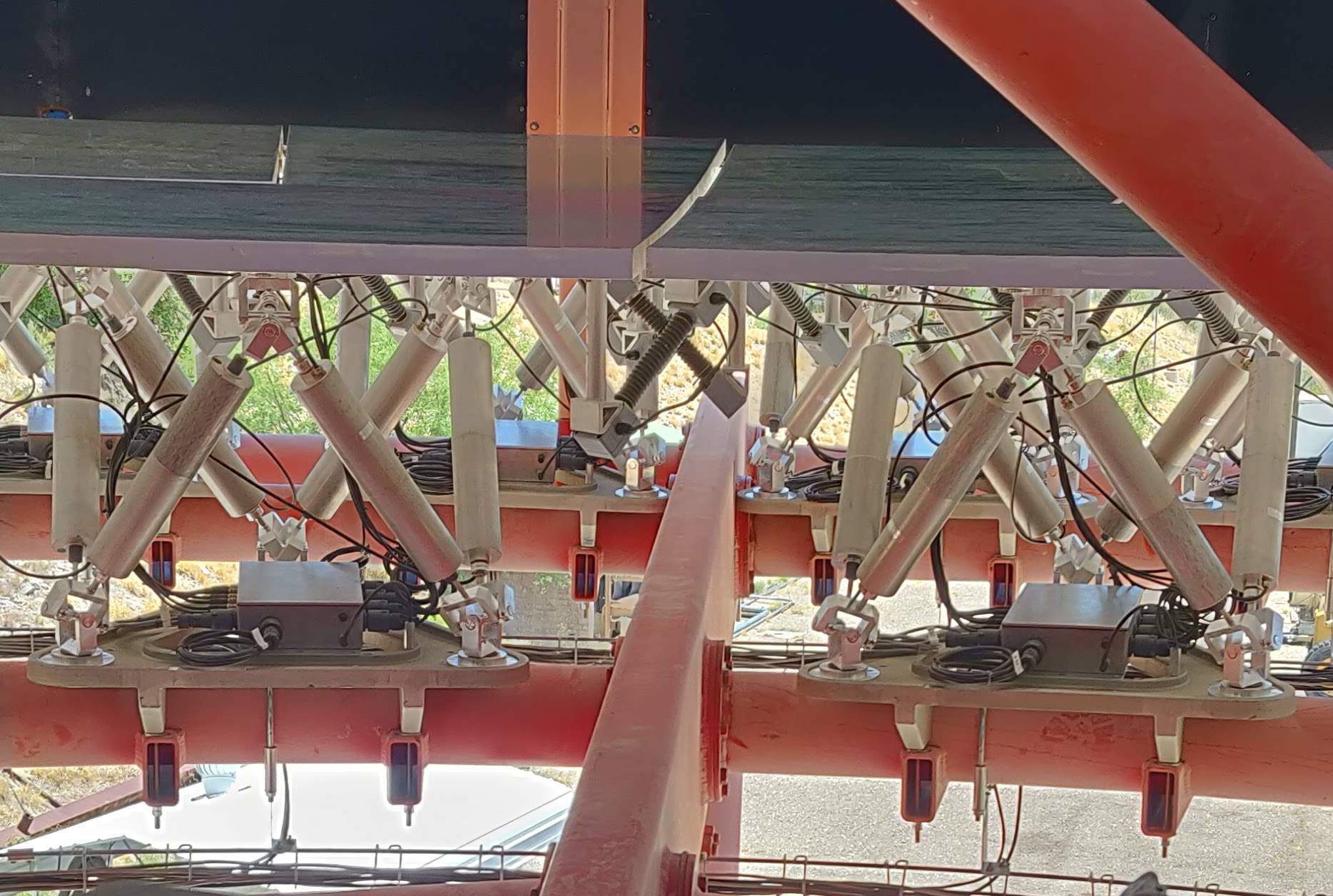}
\end{tabular}
\end{center}
\caption[GAS] 
{ \label{fig:mirror_panel_modules} 
({\it Left}) A P1 panel after assembly and MPES calibration shown during packaging for shipment to FLWO. 
The panel is lifted with a floor crane, hovering above the base of the shipping crate. The MPES units attached on the back of the mirror panel are covered for protection. Six actuators that form a Stewart platform, a controller board, and the mounting triangle interface to the OSS are visible. A custom-made blue U-shaped jig is attached to the mounting triangle and the floor crane. %
({\it Right}) Side view of two ``sectors'' of M1 panels installed on the pSCT. A P1 panel has two P2 neighbors on the outside, forming a so-called ``sector''. A ``long edge'' consisting of a P2-P2 edge and a P1-P1 edge along the azimuthal direction between two ``sectors'' is visible. 
}
\end{figure}

The pSCT OS is assembled from $72=48(\textrm{M1})+24(\textrm{M2})$ mirror panel modules (MPMs; Figure \ref{fig:mirror_panel_modules}). Each MPM is composed of a mirror panel, a Stewart platform (hexapod) that allows for the panel positioning with six degrees of freedom, several mirror-panel edge sensors (MPESs) attached on the back side of the mirror panels to measure the relative positions of neighboring panels, and a controller board with microcomputer that is responsible for collecting MPES readings, executing the re-positioning of the Stewart platform, and monitoring internal and ambient temperatures. 

\begin{figure} [ht]
\begin{center}
\begin{tabular}{c} 
\includegraphics[height=8cm]{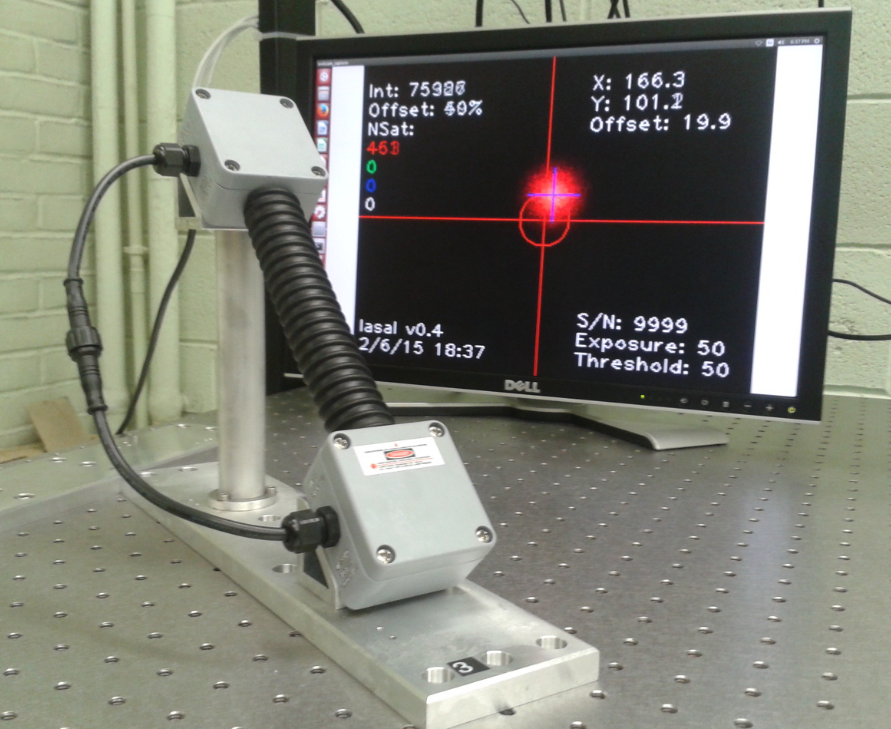}
\includegraphics[height=8cm]{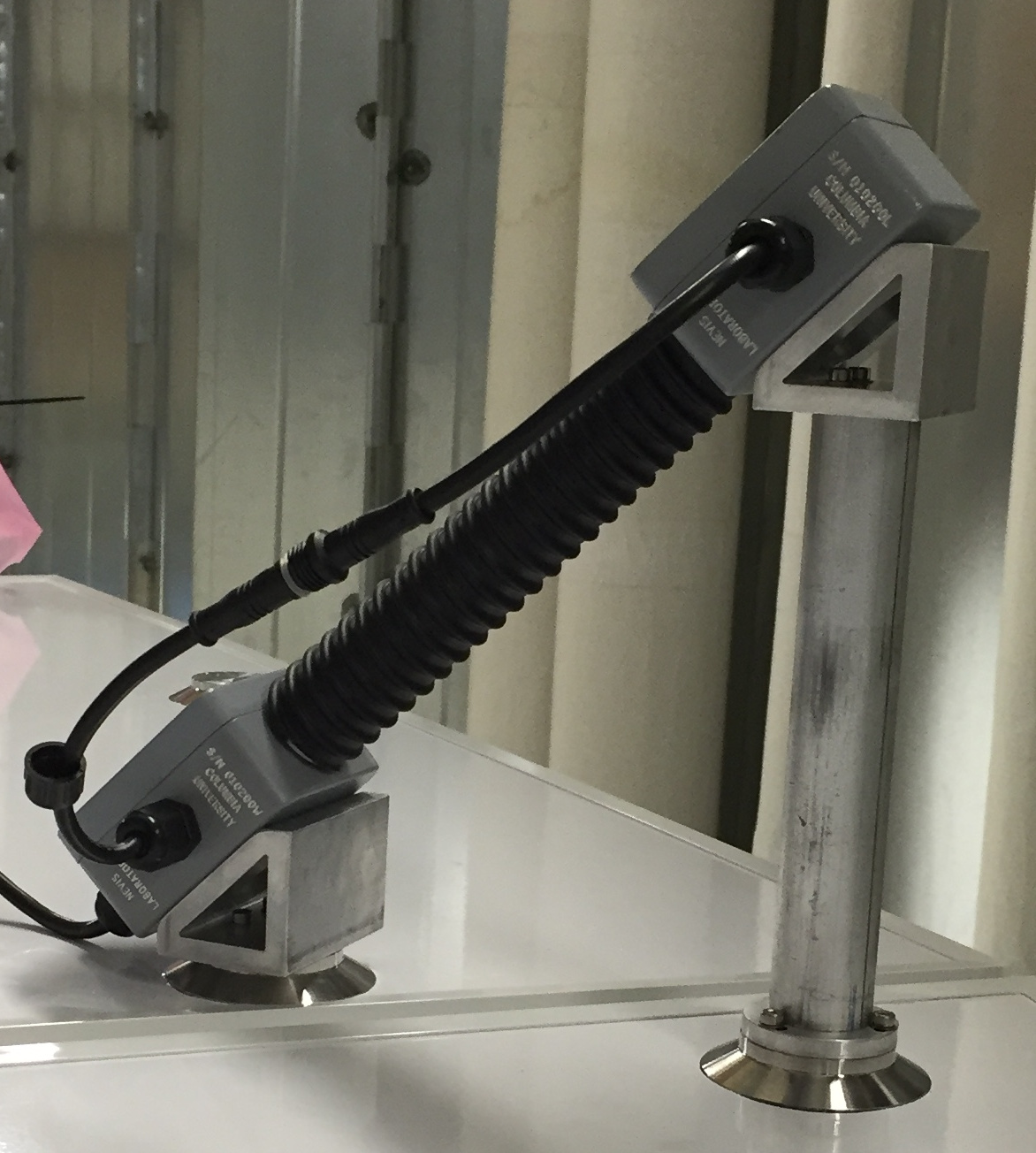}
\end{tabular}
\end{center}
\caption[GAS] 
{ \label{fig:MPES} 
The MPES is composed of a laser diode and a webcam. The webcam reading of the location of the laser beam spot, which is shown on the background monitor of the left panel in a laboratory calibration setup (this figure is taken from \citenum{Rousselle2015, Nieto15}), is used to determine the relative position of a pair of adjacent panels. The right panel shows a pair of MPES webcam and laser unit installed across the edge formed by two neighboring panels. 
}
\end{figure}

Each of the six linear actuators of the Stewart platform has
a stepping precision of 3 $\mu$m and the position of the stepping motor is measured with a magnetic encoder within the range of one revolution. Each MPES is composed of a low-power laser diode and a 0.3-megapixel low-cost webcam located on the opposite sides of an edge between a pair of adjacent mirror panels (Figure \ref{fig:MPES}). Individual MPES units are able to measure changes in the position of the laser image's centroid with a resolution better than 5 $\mu$m \citenum{Nieto15}. 

Three MPES units with mutually orthogonal laser beams are installed on each ``long'' edge  between mirror panels (between the same panel types, e.g., P1-P1) and each MPES measures the X and Y positions of the beam on the CMOS sensor of the webcam, thereby together completely constraining the relative position and orientation of one panel with respect to its neighbor. Additional MPES units installed on the edges between P1 and P2 panels and edges between S1 and S2 panels provide data for the relative positioning of the P1 and P2 rings as well as the S1 and S2 rings, and offer redundancy to mitigate the risk of MPES failures. 

While the panel-to-panel alignment system (P2PAS) based on the MPES readings monitors and corrects the relative alignment between a pair of adjacent mirror panels, a global alignment system (GAS) is designed to maintain the alignment between M1, M2 and the camera focal plane (FP). 

The GAS is assembled from two optical tables (OTs) on Stewart platforms at the centers of the primary (OT1) and secondary (OT2) mirrors, respectively and the optical camera alignment module (OCAM), which can be installed into the camera center, temporarily replacing the regular front-end electronics module of the camera located there. 

A laser mounted at the center of OT1 can produce a reference laser beam measured by transparent position sensitive devices (PSDs) installed in the OCAM and OT2. Both OTs can be actively controlled (similar to MPMs) so that the OTs and the $\gamma$-ray camera are all aligned in four degrees of freedom with respect to the reference laser beam, which then defines the optical axis.

Once the optical axis is maintained, the GAS can then monitor the relative translation and tip/tilt between the optical axis and the mirrors, both M1 and M2. The translation is measured through the analysis of images taken by three CCD cameras on each OT of three reference panels on the opposite mirror with multiple LEDs that mark the panel positions (at a 10 $\mu$m precision). 
The tip/tilt of the mirrors with respect to the OTs is measured through two autocollimators with $\sim$20 $\mu$rad accuracy, one on each OT, in combination with retro-reflectors installed on two reference panels (one on M1 and one on M2). 

OT2 is also equipped with a range meter, which allows measurement of the distance between OT1, FP and M2. Lastly, OT2 has a sky camera, the orientation of which is fixed with respect to the optical axis. The sky camera can continuously take images of the sky and provide data for offline pointing corrections of the pSCT optical axis. 

The initial alignment of the pSCT OS was performed in three steps.
The first step is the calibration of the MPMs and MPES readings in the lab, which is discussed further in Section~\ref{sec:calib}.
The second step is the MPES-guided alignment on the telescope using the reference calibrated data in the lab to achieve the rough panel-to-panel alignment, which is discussed in Section~\ref{sec:p2pas}.
The final step is the optical alignment that uses defocused on-axis star images to fine-tune alignment and optimize the PSF, which is discussed in Section~\ref{sec:opt_align}.

   \begin{figure} [ht]
   \begin{center}
   \begin{tabular}{c} 
   \includegraphics[height=8cm]{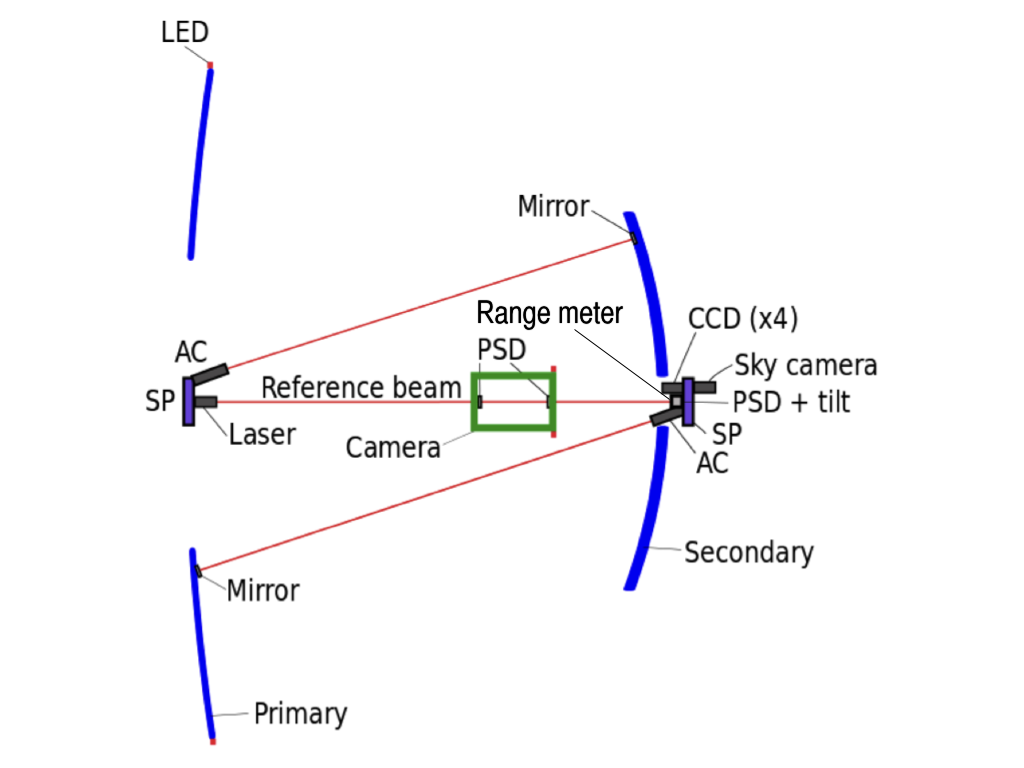}
   \end{tabular}
   \end{center}
   \caption[GAS] 
   { \label{fig:GAS} 
Illustration of the global alignment system of the pSCT, adapted from \citenum{Nieto15}. Two Stewart platforms (SPs) that support the optical tables at the center of each mirror are equipped with optical devices including autocollimators (ACs), an alignment laser, position sensitive devices (PSDs), and CCD cameras.} 
   \end{figure}

\section{Laboratory Calibration of MPES Units}
\label{sec:calib}  



As the first step of the alignment procedure, an initial laboratory calibration of the MPES units was performed following the assembly of MPMs, 
offering important information that facilitates the next step, alignment using the MPES units after the MPMs are installed on the telescope. 
The laboratory calibration process is briefly described below. 

\begin{figure} [ht]
   \begin{center}
   \begin{tabular}{c} 
   \includegraphics[height=8cm]{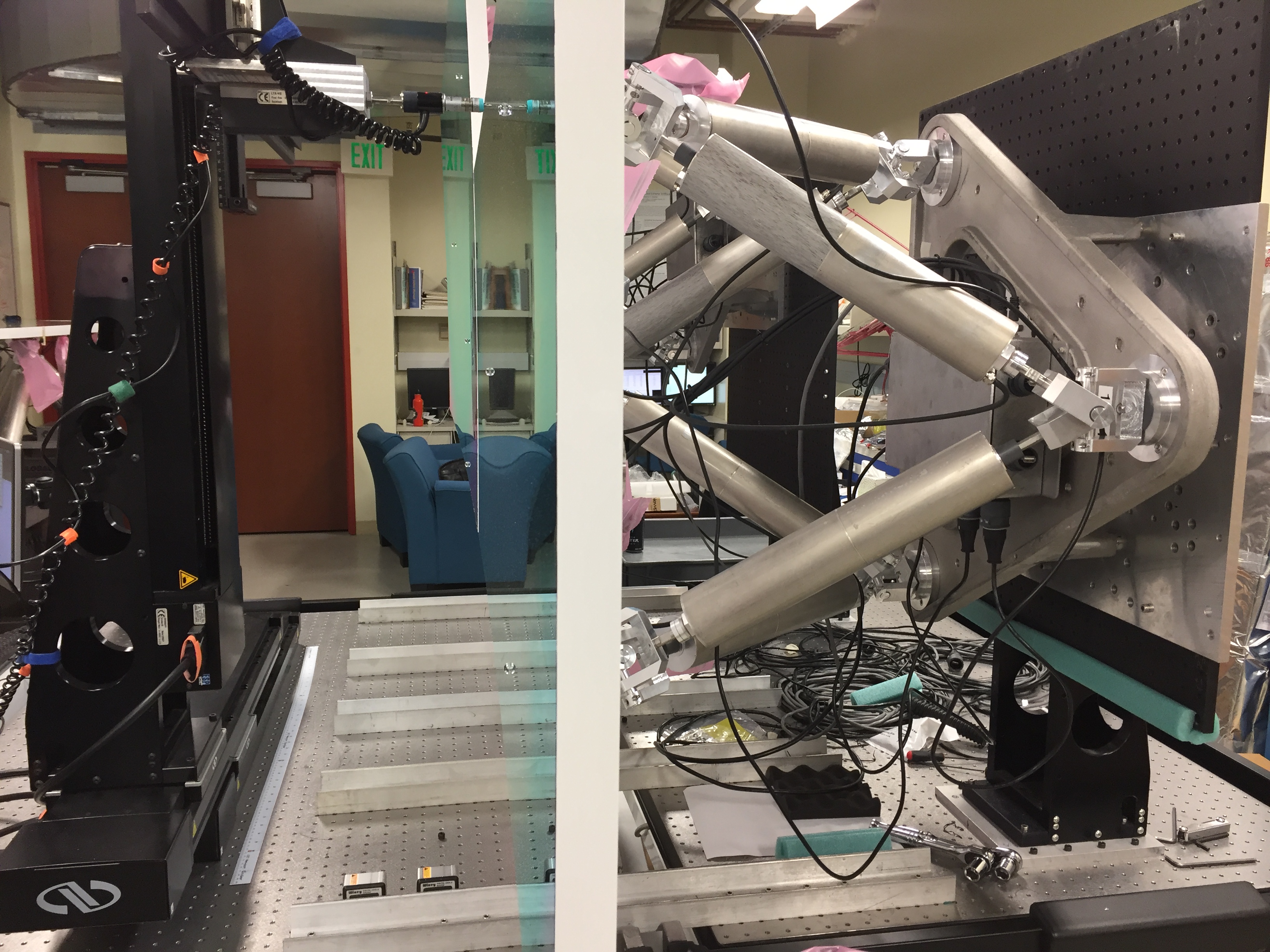}
   \end{tabular}
   \end{center}
   \caption[metro] 
   { \label{fig:metro} 
Illustration of the MPES calibration setup in the lab, with an edge-on view of two mirror panels. The white vinyl side of one mirror panel is placed vertically across the middle of the frame of this picture. The two MPMs are mounted vertically. The mounting triangle and the actuators of one panel are visible on the right-hand side of the picture. The CMM with mechanical touch probe is shown on the left-hand side of the picture.}
   \end{figure}

Each MPM was assembled by installing six actuators that connect a mounting triangle and the aluminum-coated mirror panel, forming a Stewart platform.
After two neighboring MPMs were assembled, they were mounted on a laboratory optical table, so that the edge between them could be aligned using a coordinate measuring machine (CMM) that employs a mechanical Renishaw touch probe. Figure~\ref{fig:metro} shows the calibration setup in the lab. 

After each edge was aligned, the MPES units along that edge were installed, and their readings were recorded to define the initial aligned state. 
A response matrix that maps the change in the MPES units' readings (i.e., the translation of the laser images in the MPES cameras along one edge) to the change in the actuator space (i.e., a motion introduced to the two neighboring panels along the given edge) was measured for each edge, and recorded in a database for subsequent use during the initial alignment of mirror panels on the telescope based on the MPES calibration. 
The reflectance of each mirror surface was also measured at five positions on the mirror before the MPM was crated and shipped to the FLWO site for installation. 

Note that the reference MPES readings determined by the metrology have some systematic errors associated with the CMM-based measurement and alignment. Also, the response matrices may not be accurate when the amount of misalignment is large. 
Although subject to these limitations, the laboratory calibration of the MPES units provided important initial reference positions of MPMs on the telescope and facilitated the next step, the MPES-guided alignment.

\section{MPES Guided Alignment}
\label{sec:p2pas}

Following the calibration of mirror panels with their respective actuator and MPES units, the completed panels were installed onto the telescope OSS on site for commissioning. Each panel was positioned such that every MPES had its laser spot within its camera's field of view of $12$ mm diameter, reaching a base alignment position prior to edge optimization.

\begin{figure}[h]
\centering
    \includegraphics[width=0.48\textwidth]{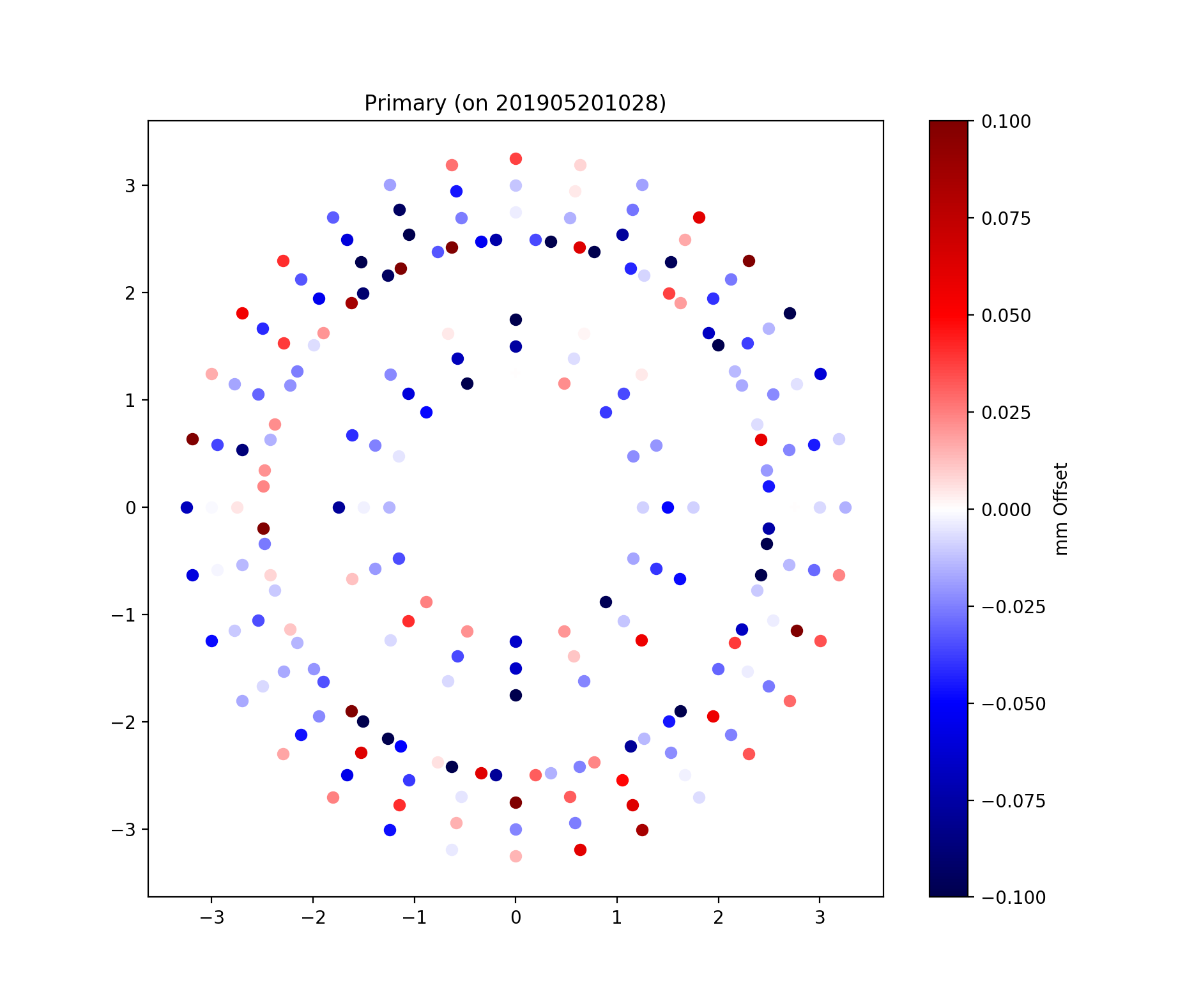}
    \includegraphics[width=0.48\textwidth]{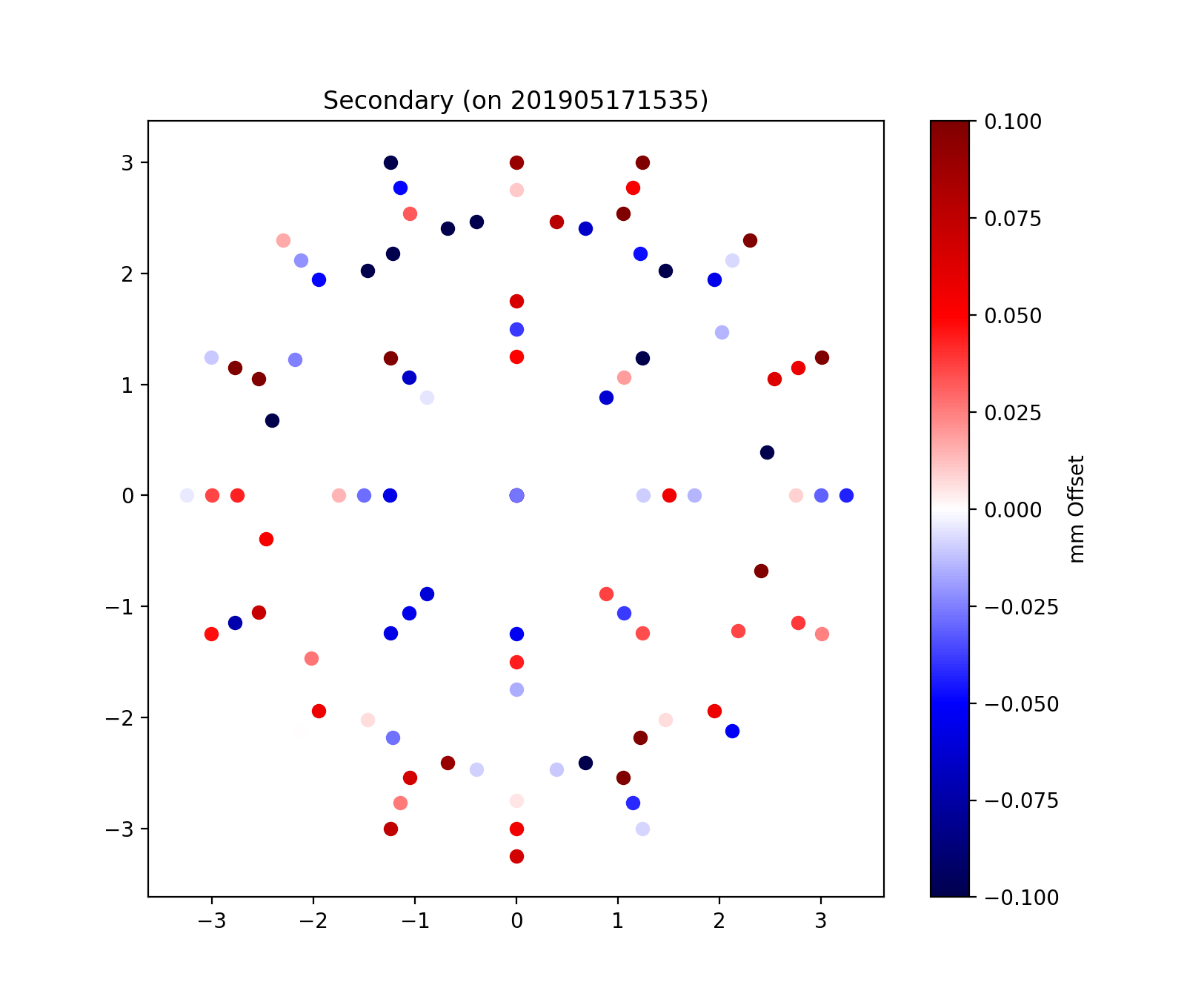}
    \caption{
    Testing of the changes in panel-to-panel alignment of the pSCT under different conditions without actively moving any optical components. ({\it left}) The differences in the panel-to-panel alignment of the primary mirrors were calculated from two sets of MPES measurements nine hours apart at temperatures of 10$^\circ$C and 20$^\circ$C, when the pSCT was parked at $20^\circ$ elevation. ({\it right}) The change in panel-to-panel alignment of the secondary mirror between 20$^\circ$ and 60$^\circ$ elevation, prior to alignment using defocused images of stars, with a mean and standard deviation of the misalignment values being 0.3~mm and 0.2~mm, respectively. Figures are taken from \citenum{Adams2019}. 
    }
\label{fig:align}
\end{figure}

With all sensors in place throughout the primary and secondary mirrors, the misalignments (relative to the lab-determined calibration) between each edge were measured and minimized iteratively with a chi-squared fitting. The panel-to-panel misalignment measurements cannot constrain well the large-spatial-scale misalignments of M1 and M2, such as dipole or quadrupole perturbation modes which at each edge between mirror panels accumulate only small distortions. The rigid-body motion of M1 and M2, for example, is not constrained by panel-to-panel mis-alignments. These loosely constrained degrees of freedom characterized by small eigenvalues of the mirror-level response matrix were regularized by introducing additional constraints such as fixed z-positions of the mirror panels with respect to the OSS, the metrology of which was previously measured. Deformations of the OSS due to elevation and temperature changes introduced initially uncharacterized misalignments throughout both M1 and M2.  
(see Figure~\ref{fig:align} for temperature and elevation drift). At this stage of alignment the structural deformations were mainly minimized by operating during stable temperatures ($<27^\circ$C) and tracking stars nearest to 60$^\circ$ elevation. The 60$^\circ$ elevation serves as a reference (bias alignment position), based on which the relative change in the OS will be studied further.  

As mentioned above, both the dipole distortion and systematic offset of M1 and M2 were decreased by utilizing regularization strategies {(to be described in detail in a separate publication)}, such as selecting particular ``reference'' panels or degrees of freedom in panel motion to minimize against. In one case, a reference panel was moved to offset dipole distortion and then frozen while the other panels were aligned using MPESs to account for this motion. In another case, inner ring panels were aligned and fixed, then the outer ring panels were brought to alignment with the inner ring while constraining their rotational degrees of freedom.

An important component of this alignment procedure was the development of software capable of parallel processing of multiple panels and/or sensors. Each component could be read out and controlled in real time, and functions were developed to control sectors of devices together, such as the creation of a ``pseudo" solid body of panels that could move in the telescope reference frame rather than individual, local panel reference frames. These methods enabled controlled motion of the entire mirror, to support large-scale corrections such as rotations of the effective mirror plane and translations to adjust the position of the focal plane.

\begin{figure}[h]
\vspace*{-\baselineskip}
\centering
    \includegraphics[width=0.49\textwidth]{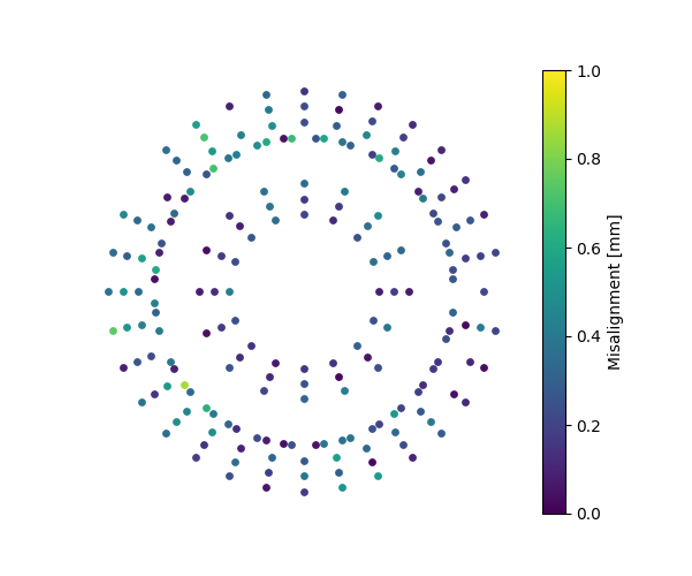}
    \includegraphics[width=0.49\textwidth]{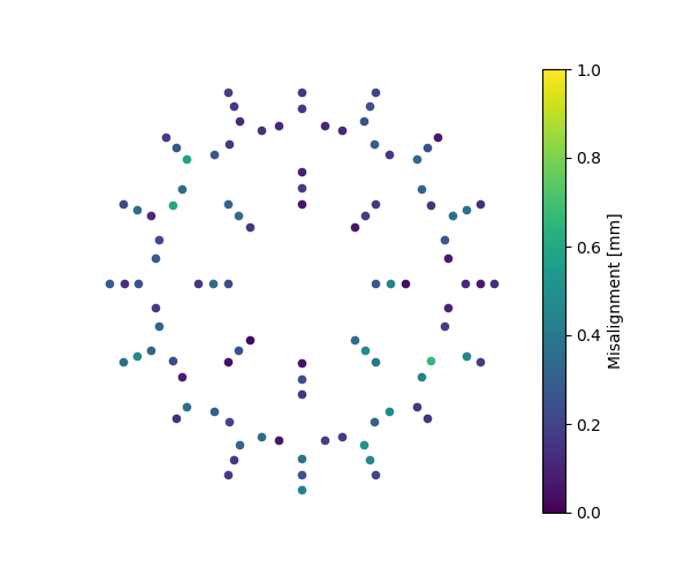}
    \includegraphics[width=0.43\textwidth]{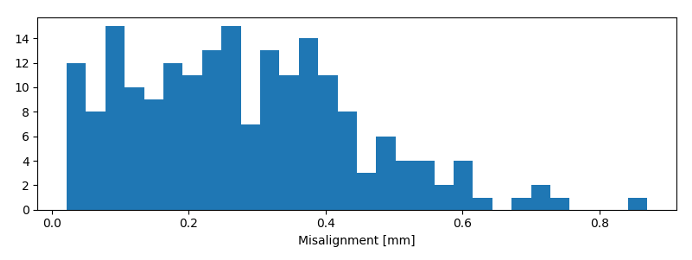}
    \includegraphics[width=0.43\textwidth]{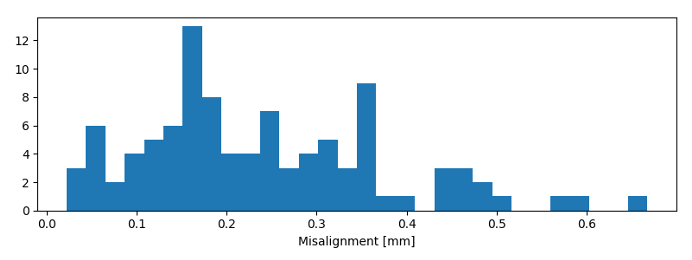}    
    \caption{
    Best alignment of M1 ({\it left}) and M2 ({\it right}) mirrors using the MPES units alone, prior to alignment using defocused images of stars. Top panels show the amount of the MPES units' displacement with respect to the laboratory-calibrated readings. The positions of the filled circles correspond to the installed locations of each MPES, and the color scale shows the amount of misalignment. The bottom panels are histograms of the displacements shown in the top panels. 
    }
\label{fig:best_alignment}
\end{figure}

The MPES-guided alignment procedure has brought the panel-to-panel alignment to a milestone shown in Figure \ref{fig:best_alignment}, where the distribution of measured offsets in MPES units averages $\approx 300$ $\mu$m. These measurement offsets are with respect to the reference MPES readings determined by the calibration in the lab, as described in Section~\ref{sec:calib}. Due to the challenge of the systematic errors described earlier in this Section, the current MPES-guided alignment results do not reflect the full capability of the MPES-guided alignment method. 
The reference MPES readings will be updated in the database replacing laboratory measurements after a precise optical alignment is achieved. The  MPES-guided alignment capability is expected to then be on the order of 50 $\mu$m or better. 

\section{Alignment Using Defocused Images of Stars}
\label{sec:opt_align}

After initial alignment using the MPES units, the optical image of a bright on-axis star was utilized to further align the mirrors. 
A misaligned panel will result in an image of the on-axis star appearing away from the focal point. By introducing a motion of the aforementioned panel so that the image of the star moves to the focal point, alignment of that panel can be achieved.  
This procedure, referred to as \emph{alignment using defocused images of stars}, offers much stronger constraints on the tip-tilt of the optical surfaces compared to alignment using MPES.

The alignment procedure using defocused images of stars is briefly described as follows: 
\begin{enumerate}

\item  Set the telescope to track a bright star at a certain elevation, the reference elevation being 60$^\circ$. Adjust the exposure of a CCD camera that captures optical images of a reflective screen placed in the focal plane so that the reflected images of the star from pairs of panels on the focal plane are visible but not saturated. 

\item Software that utilizes the \texttt{astrometry.net} tools \citenum{Lang2010} was developed for identification and characterization of the image such as centroid position, major and minor axes of light distribution, etc.. 

\item  Preparatory work to identify the image of the star from a given individual panel was done in iterations. This was accomplished by introducing a tip-tilt motion of one panel, and observing the translation of a stellar image centroid. 
Each of the 16 P1 panels reflect on-axis light onto one of S1 panels, and produce one image. Each of the 32 P2 panels reflect on-axis light onto one of S2 and one of S1 panels, and produce two images. Therefore, there is a maximum of 80 images of one on-axis star, if all panels are misaligned. 

\item Complex optical coupling of the images described can be efficiently utilized to align the pSCT OS. If the S1 and S2 panels are misaligned with respect to each other, a P2 panel will produce two images of an on-axis star. This can be tested by observing a translation of a pair of stellar images on the focal plane when a single P2 panel is tipped/tilted. 
Therefore, the alignment between the S1 and S2 panels can be done by applying a tipping motion (a change in pitch angle along the azimuthal axis of the ring of mirrors) to a S2 panel, the amount of which corresponds to the offset between the image pair from the same P2 panel. 

\item  A response matrix that maps the translation of an image centroid of the star to the tip-tilt motion of a corresponding panel was measured for each panel. 
These matrices were then used to calculate the tip-tilt motion that can translate the stellar image by any necessary amount. Note that the implied linear approximation means that the alignment using defocused images of stars is most accurate when misalignment is small. 

\item  Once all pairs of panels and optical images were identified and response matrices measured, tip-tilt motions of all primary panels that can translate all optical images to the focal point were calculated and executed. 
If the initial misalignment is large, iterations are needed to improve alignment due to the limits of assuming a linear approximation. 
Because the alignment using defocused images of stars does not provide a strong constraint on the translation of individual panels, alignment using the MPES units are performed to constrain panel translation and maintain an overall panel-to-panel alignment in the process. 

\item  A ``defocused'' pattern, in which optical images from individual primary panels are intentionally placed into ring patterns (mimicking the layout of the panels for easy identification of images and efficient alignment using defocused images of stars), is established, and enables rapid parallel measurements of all response matrices. This is especially useful for monitoring the effects of OSS deformations under elevation-dependent gravitational load and temperature variations, as well as for repetition of the alignment procedure using defocused images of stars at different elevation and azimuth angles. 

\end{enumerate}

\begin{figure}[h]
\vspace*{-\baselineskip}
\centering
    \includegraphics[width=0.47\textwidth]{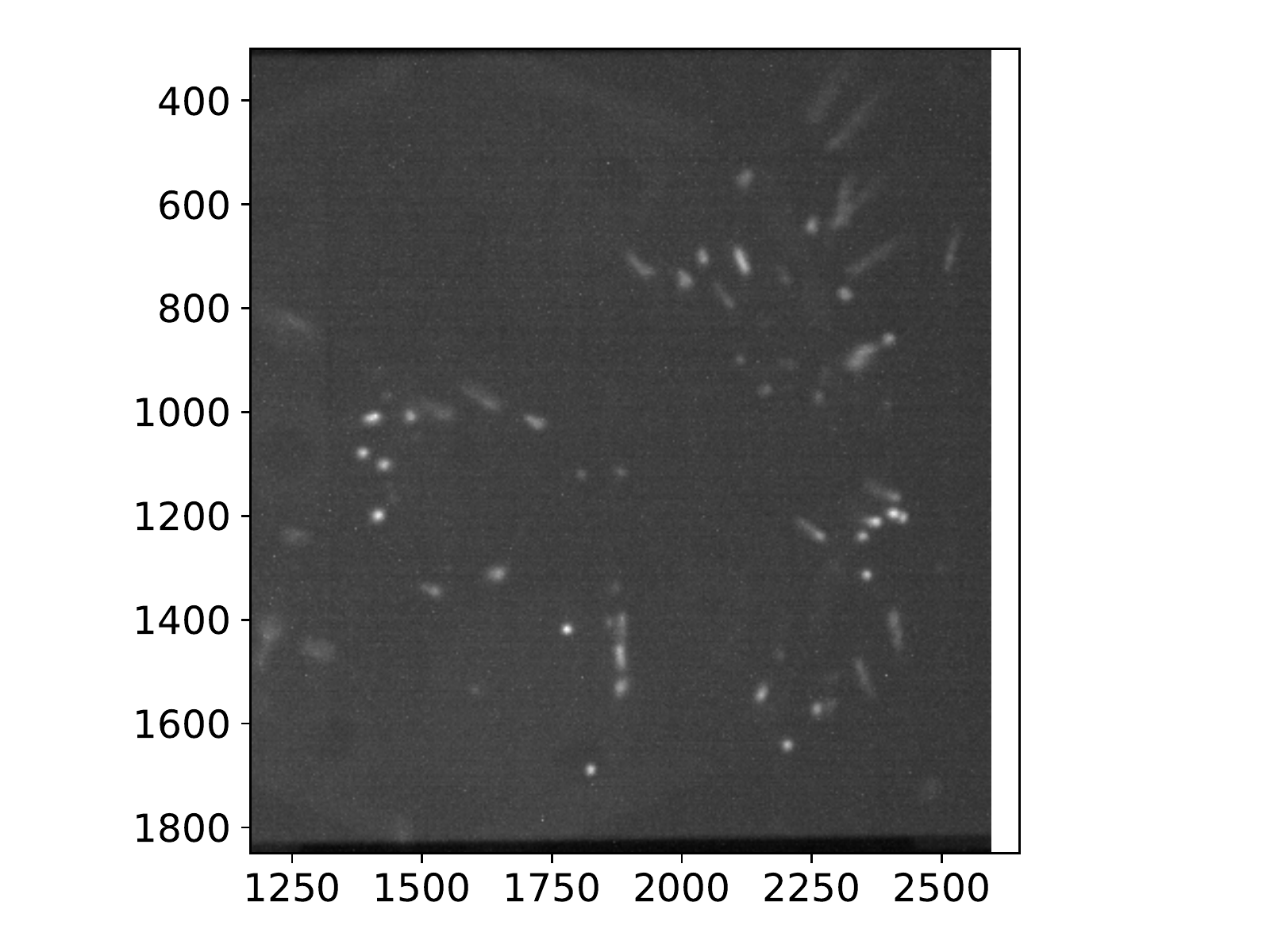}
    \includegraphics[width=0.47\textwidth]{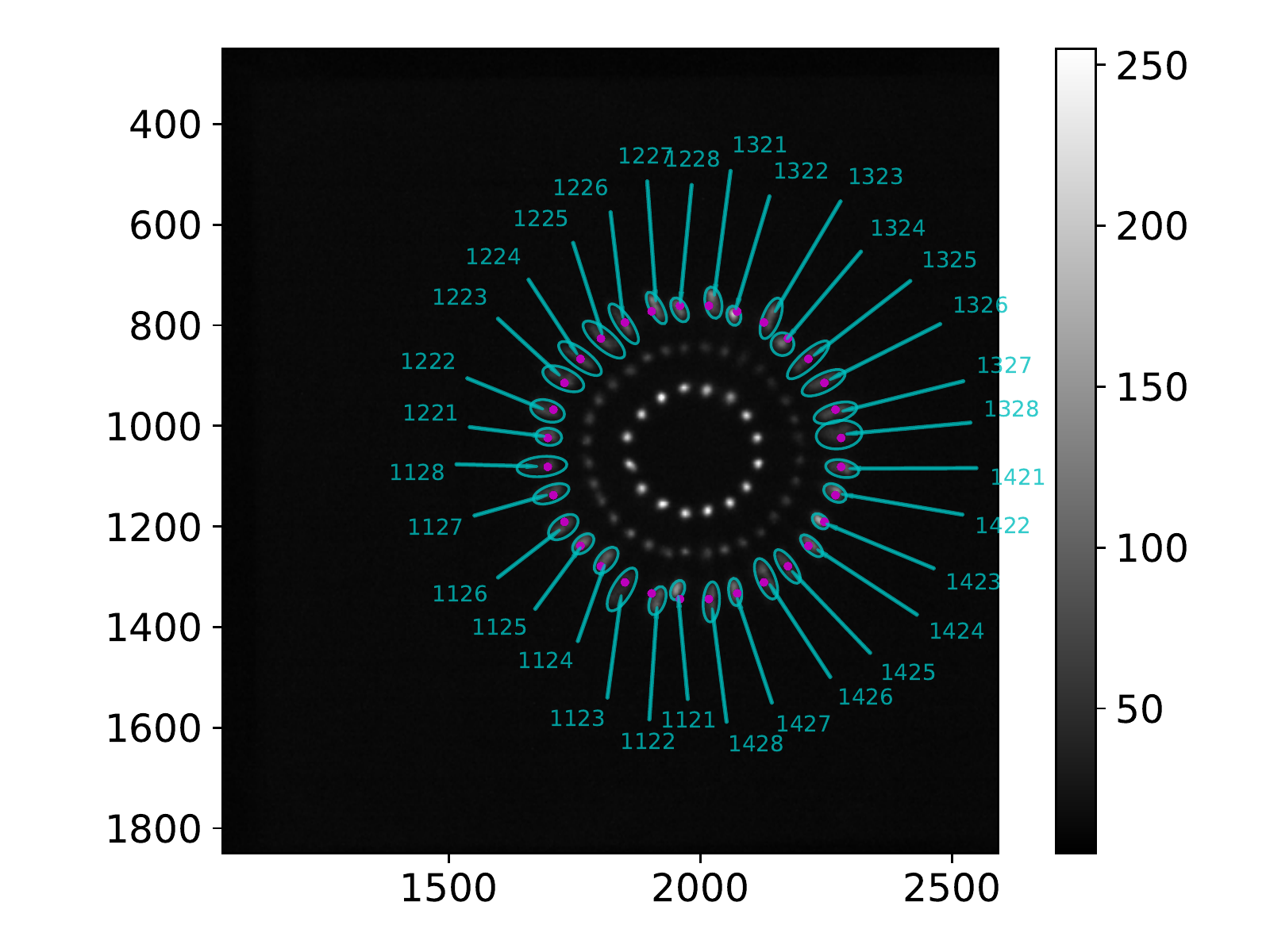}
    \caption{Optical alignment intermediate steps: ({\it left}) Image of a bright star following the initial alignment based solely on the MPES units. ({\it right}) Defocused optical images of a bright star with P2 panels identified. Both images shown cover about a 3.5$^\circ$ field with the focal point and the pSCT camera roughly in the center. 
    }
\label{fig:optical_start_middle}
\end{figure}

\begin{figure}[h]
\vspace*{-\baselineskip}
\centering
\includegraphics[width=0.47\textwidth]{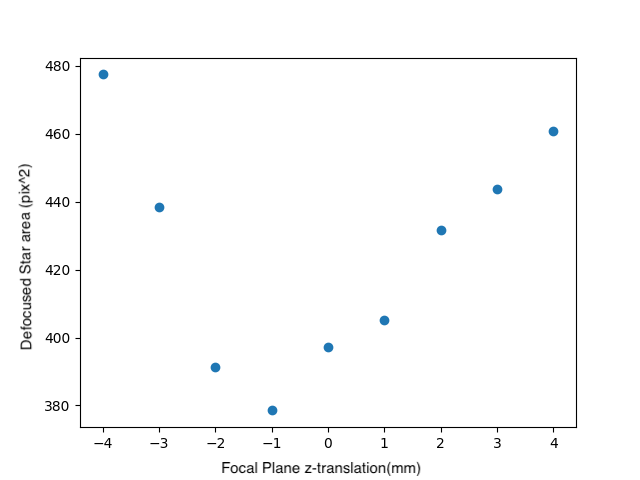}
\includegraphics[width=0.47\textwidth]{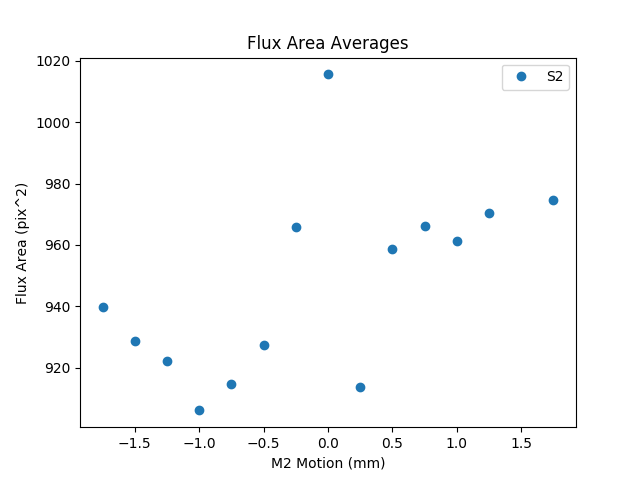}
\caption{
    The average area ($\sim$80\% containment) of individual images of stars on the focal plane produced by P1-S1 panels ({\it left}) and P2-S2 panels ({\it right}) with the ``defocused'' configuration (as shown in the right panel of Figure~\ref{fig:optical_start_middle}) with respect to the distance between the $\gamma$-ray camera and M2.
    The defocused stellar images from P1-S1 form the inner circle shown in Figure~\ref{fig:optical_start_middle}, and those from P2-S2 form the outer circle. 
    Each data point is derived from an image similar to Figure~\ref{fig:optical_start_middle} when the distance between $\gamma$-ray camera and M2 is at one of the nine relative values in 1-mm increments shown in the plots.
    At the ``-1 mm'' relative distance between M2 and the $\gamma$-ray camera, the defocused images of the stars reach a minimal area, and the OS is the closest to true focus. 
    }
\label{fig:optical_FocalPlane_height}
\end{figure}


The left panel of Figure~\ref{fig:optical_start_middle} shows an optical image of an on-axis star after the initial alignment using only MPES units. 
Despite the sub-mm misalignment of mirror panel edges achieved with the MPESs alignment, these images scatter across and beyond the $\sim3.5^\circ$ field that is shown given the $\approx 10$ m distance between M1 and FP.

The right panel of Figure~\ref{fig:optical_start_middle} shows a pattern of defocused optical images of an on-axis star, after initial optical alignment was achieved, following the steps described above. In this configuration, the S1 panels remain in the aligned orientation; all primary panels and the S2 panels are tipped outward, i.e., the pitch angle along the azimuthal axis of the ring of mirrors changed so that the ring becomes ``flatter''. The 16 optical images that form the inner circle in the right panel of Figure~\ref{fig:optical_start_middle} correspond to the images produced by pairs of P1 and S1 panels, and can be uniquely and quickly identified in software. Similarly, the images that form the outer circle are from pairs of P2 and S2 panels, and the corresponding P2 panel identification numbers are shown in cyan. Lastly, the faint images that form the middle circle are from pairs of P2 and S1 panels. The lower light intensity from the middle circle compared to the outer circle is because less light from the P2 panels is reflected from the S1 panels than from the S2 panels. 
When the S1 and S2 panels are aligned with respect to each other, the outer and middle circles merge into one. 

Another important quantity of the optical images of on-axis stars is the image shape. Elongation of the optical images occurs as they move away tangentially from the focal point as panels tip or tilt. The area of the optical images increases as the distance between the $\gamma$-ray camera and the secondary mirror deviates from the focal length. Therefore, the elongation and the area of the optical images were used to correct for any global offset (both rotations and translations) between the $\gamma$-ray camera FP and the M2. 
Specifically, a global tip-tilt motion of the entire secondary mirror from the ``defocused'' configuration can be applied until the ellipticities of all the images are the same, corresponding to no relative tip-tilt between the $\gamma$-ray camera and the secondary mirror. 

The PSF of the SC OS is relatively insensitive to the position of the M1 mirror relative to M2. However, it is highly sensitive to the relative placement of the M2 mirror and the focal plane which by design are separated by $1.86$ m as shown in Table~\ref{tab:telescope}.
The initial focus of the SC OS was achieved by changing the position of the FP in the OS and hence the distance between the $\gamma$-ray camera and M2, and minimizing the area of the optical images of defocused stars. Figure~\ref{fig:optical_FocalPlane_height} shows this first, low-resolution,  attempt to focus the pSCT by adjusting the distance between the $\gamma$-ray camera and M2 with a 1-mm precision.
At the ``-1 mm'' relative distance between M2 and the $\gamma$-ray camera, the system is the closest to true focus. A smaller increment of $0.25$ mm in the distance adjustment between FP and M2 was used to further improve the focus of the OS by moving M2 as a whole. No effort was made at this stage to optimize the position of M1 and it is currently placed at the default position shown in Table~\ref{tab:telescope}.

Alignment using defocused images of stars is one of several possible alignment methods that utilizes the optical surface, and can be used to effectively constrain a number of degrees of freedom in the alignment of the pSCT OS. Future alignment work that further exploits the benefits of this method will be carried out. 



\section{Optical PSF}
\label{sec:psf-res}
\begin{figure} [ht]
   \begin{center}
   \begin{tabular}{c} 
   \includegraphics[height=8cm]{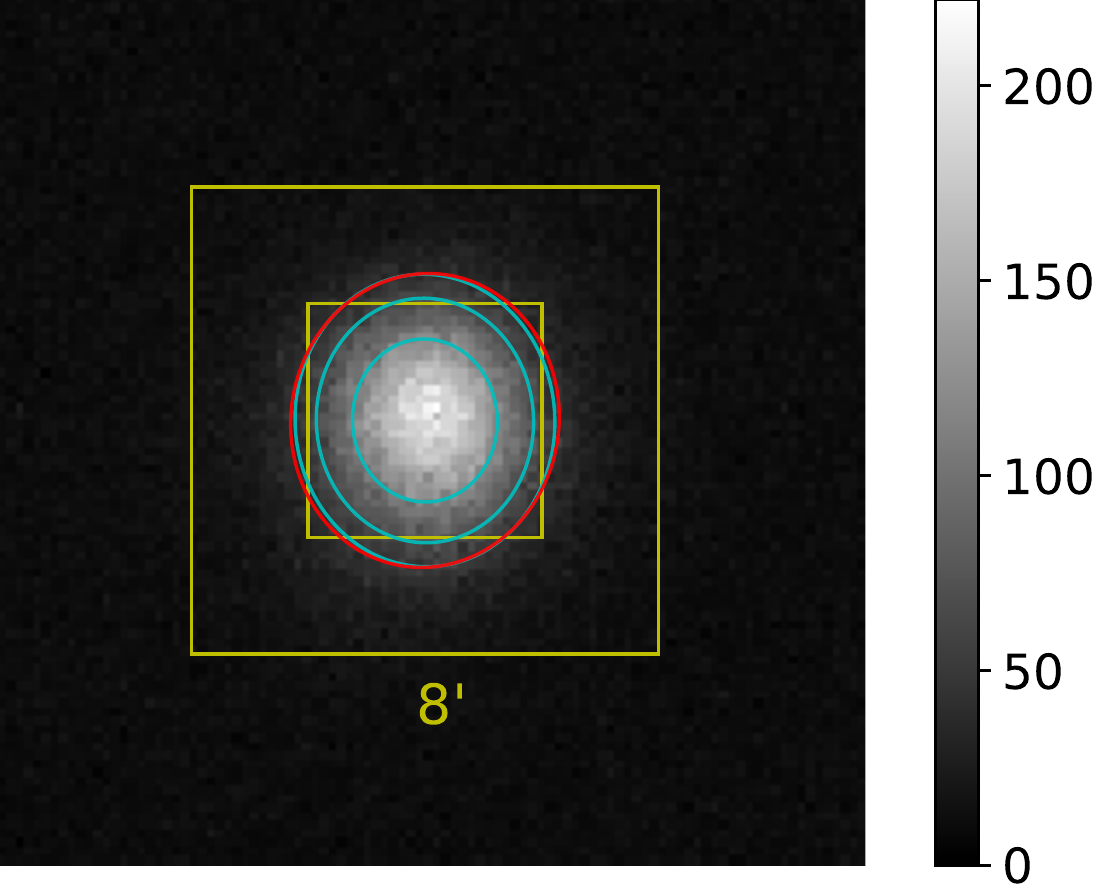}
   \end{tabular}
   \end{center}
   \caption[PSF] 
   { \label{fig:psf_zoom} 
The PSF of the pSCT, illustrated by an image of the star Capella on the focal plane, as achieved during the initial commissioning campaign concluded in December 2019. From inner to outer, the three cyan ellipses show the 1-$\sigma$ ($\sim$39\% containment), 1.5-$\sigma$ ($\sim$68\% containment), and 1.8-$\sigma$ ($\sim$80\% containment) contours from the best 2D Gaussian fit, respectively. The red contour is the extension solution obtained from the \texttt{astrometry.net} tool, roughly corresponding to $80$\% light containment without reliance on the 2D Gaussian fit. The inner and outer yellow unfilled squares illustrate the sizes of the pSCT “imaging” pixel (an angular scale of 4 arcmin) and the pSCT “trigger” pixel (8 arcmin), respectively.}
   \end{figure}

The PSF requirement and goal specifications for the pSCT were defined through detailed simulation studies of the PSF error budget and were reported during the pSCT pre-construction readiness review conducted by the CTA project on September 28, 2013.  Following the definition of the optical PSF in this study, PSF=$2\times \max \{\sigma_x, \sigma_y\}$ 
(where $\sigma_x$ and $\sigma_y$ are root mean square values along two orthogonal axes of the PSF),
an on-axis PSF of $3.6$ arcmin was defined as  “acceptable” and an on-axis PSF of $2.6$ arcmin was defined as the “goal”. During the phase of initial commissioning of the pSCT in December 2019, an on-axis PSF of $2.8$ arcmin has been achieved after following the alignment procedures described in the preceding sections, starting with the initially MPES-based alignment, which was followed by a few iterations of alignment based on the defocused images of stars, as shown in Figure~\ref{fig:psf_zoom}. 

To evaluate the pSCT PSF, the CCD image of an on-axis star was fit with a 2D Gaussian model, with seven parameters: two centroid coordinates, two centroid standard deviations ($\sigma_x$ and $\sigma_y$) along major and minor axes, respectively, a normalization constant, a major-axis orientation angle, and a constant baseline that describes the ambient light and noise associated with the CCD camera. It was verified that the constant baseline, about $6$\% of the peak intensity shown in this figure, has a negligible radial gradient and, therefore, is not associated with the stellar image itself. This fit yielded standard deviations of $\sigma_x = 1.4$ arcmin and $\sigma_y = 1.24$ arcmin, corresponding to half of the $39$\% containment major and minor axes of ellipse, respectively. It was also confirmed that  the root mean square (the second order moments) values are in agreement with the standard deviations. 

The $68$\% containment ($\sim$1.5$\sigma$) major and minor axes are $4.2$ arcmin ($6.8$ mm) and $3.8$ arcmin ($6.2$ mm), respectively, and those for 80\% containment ($\sim$1.8$\sigma$) are $5$ arcmin and $4.5$ arcmin. As illustrated by the yellow squares in Figure~\ref{fig:psf_zoom}, 75.5\% of the PSF is contained within a $4$-arcmin ($6.5$ mm) square imaging SiPM pixel centered at the peak, and 99.5\% of the PSF is contained within a 8-arcmin “trigger” pixel defined as $2 \times 2$ imaging pixels for the pSCT. Given these results, the on-axis performance of the pSCT OS is considered verified for the first time in a telescope with a 9.7-m aperture with both primary and secondary mirrors segmented. This achievement confirms that the SCT is a viable contender for the medium-sized telescope of CTA, demonstrating the advantages of the Schwarzschild-Couder OS for ground-based $\gamma$-ray astronomy applications and proves that the currently existing technologies for the manufacturing and aligning of the SC OS can be implemented at a competitive cost. 

The commissioning of the pSCT OS was interrupted by the development of the COVID-19 pandemic in the U.S. and the temporary closing of FLWO, which has not yet allowed final assessment of the optimal OS performance. Further improvement of the optical alignment is foreseen by introducing second-order corrections into the predicted tip/tilt motion of each mirror panel to realize the OS focus and by adjusting the positions of each mirror panel to correct manufacturing errors in the figures of the optical surfaces resulting in a finite spread of their focal lengths. Hence, the achievement of the on-axis PSF “goal” specification is assessed as credible at this stage.

The off-axis performance of the pSCT OS has not been fully examined yet and only an initial verification has been performed. It was confirmed that the off-axis defocused images of bright stars at $1.5^{\circ}$ field angle have not suffered noticeable degradation. The pSCT bending model and pointing corrections have not yet been implemented. In spite of these temporary limitations which will be remedied in further commissioning activities, the pSCT has demonstrated its end-to-end technology validation by the detection of the Crab Nebula\footnote{\href{https://aas.org/sites/default/files/2020-06/vandenbroucke_aas236.pdf}{https://aas.org/sites/default/files/2020-06/vandenbroucke\_aas236.pdf}}, a standard $\gamma$-ray source for the VHE astronomy.

\acknowledgments 
 
We gratefully acknowledge support from the agencies and organizations listed under Funding Agencies at this
website: \href{www.cta-observatory.org}{www.cta-observatory.org}. The development of the prototype SCT has been made possible by funding provided through the NSF-MRI program under awards PHY-1229792, PHY-1229205, and PHY-1229654. The authors are also grateful for the generous support of the National Science Foundation under awards PHY-1607491, PHY-1352567, PHY-1505811, PHY-1806554, and PHY-1913798.
We are grateful to INAF for having supported part of the M2 production. 

\bibliography{ref} 
\bibliographystyle{spiebib} 

\end{document}